\documentclass[12pt]{article}
\input epsf
\usepackage{amssymb}
\usepackage{amsmath}
\makeatletter
\@addtoreset{equation}{section}
\makeatother

\def\be{\begin{equation}}
\def\ee{\end{equation}}

\def\ba{\begin{array}}
\def\ea{\end{array}}

\newcommand{\bea}{\begin{eqnarray}}
\newcommand{\eea}{\end{eqnarray}}
\def\N{$\cal N$}

\begin{document}
\hfill{}

\vskip 2cm

\vspace{24pt}

\begin{center}
{ \LARGE {\bf  The Effective Action of  \N=8 Supergravity  }}

\vspace{24pt}

  {\large  {\bf   Renata Kallosh}}

    \vspace{15pt}

 {Department of Physics,
    Stanford University, Stanford, CA 94305}

\vspace{10pt}

\vspace{24pt}

\end{center}

\begin{abstract}

We present a simple form of the on-shell gauge-invariant 1-loop effective action of \N=8 supergravity which is manifestly  \N=8 supersymmetric at the linear level. By generalizing the dimensional arguments in superspace to non-local invariants, we show that the 1-loop effective action does not contain any contributions from  bubble and triangle diagrams. The absence of bubbles implies the absence of  conformal and axial anomalies.  We also show that the 1-loop effective action of \N=8 supergravity features a ``dual'' conformal symmetry in the momentum space.

\end{abstract}
\newpage

\section{Introduction}

There is a chance that \N=8 supergravity is perturbatively finite, according to \cite{Bern:2007hh,Bern:2006kd}. If this is the case, we will have an excellent  model of quantum gravity in four dimensions.  It was suggested in \cite{Bern:2007xj} that the finiteness may be intimately linked to the surprising cancellations
observed in the complex high energy behavior of tree amplitudes noted in
the construction of on-shell recursion relations.

The results of the 3-loop computations have been reported in \cite{Bern:2007hh}. The 3-loop  answer has several striking features.
In particular, there are no local UV divergent term of the form  $  (R_{....})^4$. This happens not because of the cancellation of different terms: the UV divergent terms never even show up in any diagram in  \cite{Bern:2007hh}.
There are no non-local terms of the form  $   \partial ^2 (R_{....})^4$, these terms also never show up.
There are no non-local terms of the form  $  \partial ^4 (R_{....})^4$.
This  is a result of the highly non-trivial cancellation  of such terms which are present in quite a few diagrams.
The first non-local terms which are present are related to $  \partial ^6 (R_{....})^4$.
This behavior was predicted on the basis of the so-called ``no-triangle hypothesis in \N=8 supergravity'' in \cite{BjerrumBohr:2006yw}. This hypothesis at the level of 3-loops, as well as its prediction, were  confirmed by explicit 3-loop computations.

On the other hand, many supergravity experts still believe  that starting from some loop order (depending on various assumptions) there may be an infinite number of divergences.   Currently, there is no agreement concerning the onset of UV divergences.   According to \cite{Howe:1980th}, \cite{Kallosh:1980fi},
they may start at the 8-loop level, or even earlier, at the 3-loop level, according to \cite{Kallosh:1980fi,Howe:1981xy}, or from the 5-loop level \cite{Howe:2002ui}. The reason why it may be difficult to convince the supergravity community that \N=8 is all-loop finite is the following. The lore about UV infinities in gravity was established a long time ago, when it was proposed, on the basis of  local counterterm analysis in the framework of the background  method in background covariant gauges, that pure gravity may be UV divergent starting from the 2-loop level.  It was shown that only one of the  many  {\it a priori} available invariants  can describe the gauge-independent on shell UV divergence \cite{Kallosh:1974yh,vanNieuwenhuizen:1976vb}, namely
\be
\Gamma_{2} \sim {1\over \epsilon} \kappa^2 \int d^4 x \sqrt {-g}  \, C_{ab}{}^{cd} C_{cd}{}^{ef}   C_{ef}{}^{ab} \ .
\label{prediction}
 \ee
Here $C_{abcd}$ is the conformal Weyl curvature tensor, and an on-shell condition is
$
R=R_{ab}=0$ and $ C_{abcd} = R_{abcd}
$.
There was an extensive verification of the background field methods in background covariant gauges in gravity and Yang Mills theory, in particular in \cite{Grisaru:1975ei}.   12 years after the prediction (\ref{prediction}), the actual computation was performed  in \cite{Goroff:1985th}, which demonstrated that 2-loop gravity is UV divergent:
\be
\Gamma_2=  {\kappa^2\over (4\pi)^4 } {209\over  2880} \, {1\over \epsilon}\,  \int d^4 x \sqrt {-g}  \, C_{ab}{}^{cd} C_{cd}{}^{ef}   C_{ef}{}^{ab} \ .
\label{prediction1}\ee
 6 years later an analogous  computation was performed in
\cite{vandeVen:1991gw},  in more sophisticated  non-linear background covariant gauges. The numerical answer was the same. This finalized a convincing story of the UV infinities in quantum gravity. There was no reason to expect that the 3-loop  counterterm
\be
S_{3} \sim {1\over \epsilon} \kappa^4 \int d^4 x \sqrt {-g}  (C_{....})^4 \ ,
\label{predictionS} \ee
as well as all higher loop order  ${1\over \epsilon}$ UV divergences, would not show up.

With this mindset, the UV divergences in supergravity were analyzed. Every pure supergravity was shown to be free of  2-loop divergences, but  the candidate for the 3-loop divergence, a supersymmetric generalization of the square of the Bell-Robinson tensor, was proposed in \cite{Deser:1977nt}.

In \N=8 supergravity, the \N=8 supersymmetric extension of the  local counterterm in the 3d loop (\ref{predictionS}) was proposed in \cite{Kallosh:1980fi,Howe:1981xy}.  It was also expected that all ${1\over \epsilon}$ terms will be local counterterms of the kind given in  (\ref{predictionS}),  just supersymmetrized. For example, starting from the  8-loop order   non-linear, geometric \N=8 invariant counterterms were  constructed \cite{Howe:1980th}, \cite{Kallosh:1980fi} using the \N=8 on shell superspace \cite{Brink:1979nt}. The first full superspace integral over torsions, a candidate for the 8-loop UV divergence,  is given by  \cite{Howe:1980th}, \cite{Kallosh:1980fi},
\be
S^8 \sim \kappa^{14} \int d^4 x \; d^{32} \theta  \; \rm Ber E \; T_{ijk \alpha} (x, \theta)  \overline T^{ijk \dot \alpha} (x, \theta)  T_{mnl }{}^{ \alpha} (x, \theta) \overline T^{mnl }{}_{\dot \alpha}(x, \theta)  \ .
\label{S8}\ee
Here $T_{ijk \alpha}(x, \theta) $ is the superspace torsion superfield whose first component is a  spinor  field. The scalars of \N=8 supergravity are in the  ${E_{(7,7)}\over SU(8)}$ coset space and the spinor superfield is related to the spinorial derivative of the scalar superfield.

The geometric superinvariants like the one in eq. (\ref{S8}) have unbroken hidden  $E_{(7,7)}$  symmetry: the torsion tensor is a  $E_{(7,7)}$ invariant, and an $SU(8)$ tensor. This symmetry  plays an important role for the black holes in \N=8 supergravity \cite{Kallosh:1996uy} since the entropy is given by the quartic invariant of $E_{(7,7)}$.

The bosonic part of an infinite number of UV counterterms quartic in Weyl tensor for arbitrary loop order $L$ is of the general form
\be
S^{L}\sim \kappa^{2(L-1)} \int d^4 x  C^2(x) \partial ^{L+2} C^2 +...\ , \qquad L\geq 8 \ .
\ee
These terms can be presented in a manifestly \N=8 supersymmetric form  with the hidden $E_{(7,7)}$ symmetry. And, of course, many other geometric invariants with higher powers of superfields are possible.

The existence of counterterms is not the  full story of the  attempts to understand the UV properties of supergravities and super Yang Mills theories. A careful treatment of the supergraphs in the background field method in  supersymmetric theories was made in \cite{Grisaru:1979wc}, \cite{Howe:1983sr} and in the  harmonic superspace in \cite{Galperin:2001uw}, \cite{Hartwell:1994rp}. It become clear that
 counterterms need to be writable in terms of full superspace integrals for the linearly realizable off-shell supersymmetry, at least in the N=4 super-Yang-Mills case.  The analogous issue in supergravity is still to be clarified.

Thus, the computational precedent in pure gravity, as well as the existence of local  \N=8 supersymmetric $E_{(7,7)}$ invariant UV counterterms,  is the basic reason why there is an expectation in the supergravity community that \N=8 supergravity is likely not to be UV finite, at least starting from the 8-loop order.  It is also not clear whether one can trust the finiteness arguments for  \N=8 d=4 supergravity derived from string theory,  because of nondecoupling of the non-perturbative states
\cite{Green:2007zz}.

It is therefore important either to  actually prove the finiteness of  \N=8 supergravity to all orders of perturbation theory, as proposed in \cite{Bern:2007hh,Bern:2006kd}, or prove the opposite. One may try to use the available supergravity tools for this purpose.  These tools are based on superfields in  manifestly realized supersymmetry in Yang Mills theory, \cite{Ferrara:1974pu} and in supergravity \cite{Ferrara:1977mv} and their generalization to higher \N.

In this note we  use the  background field method and manifestly supersymmetric on-shell superfields of  \N=8 supergravity, in addition to inferring the information from either \N=4 Yang-Mills theory, or from string theory \cite{Bern:2007hh,Bern:2006kd}.
One should keep in mind that in the background effective action method we deal only with one-particle irreducible graphs: the background fields satisfy a non-linear equation of motion which allows one to include also the trees. The translation between the S-matrix  helicity amplitudes and the gauge invariant background field method will require extra work, but it may be justified.

The Feynman rules of \N=1 supergravity have been used in \cite{Grisaru:1981ye}, \cite{Gates:1983nr} to compute the 1-loop four-particle S-matrix in
\N=8 supergravity. Their result via supergraphs  agrees with an  earlier string theory computation in the   d=4 limit \cite{Green:1982sw}.

We will  generalize  the answer for the effective action obtained from \N=1 supergraph rules  in  \cite{Grisaru:1981ye}  so that it becomes  \N=8 supersymmetric.  The main idea behind such a supersymmetrization is to use the building blocks constructed  in the past in \cite{Kallosh:1980fi,Howe:1981xy} for the higher loop local UV counterterms and apply them to the 1-loop nonlocal invariants.

 This will bring us naturally to the issue of conformal and axial anomalies in \N=8 supergravity which was studied extensively in the past \cite{Grisaru:1981bs}, \cite{Duff:1982yw}, \cite{Kallosh:1980fi},  \cite{Gates:1983nr}. At that time, the status of anomalies remained inconclusive. We will see here that the 1-loop ``no-triangle'' hypothesis proposed in \cite{BjerrumBohr:2006yw},  which provides the basis for the conjectured all-loop finiteness of \N=8 supergravity,   can be related to the absence of the 1-loop anomalies.

As a bonus of constructing the 1-loop effective action  of \N=8 supergravity, we will find a connection to the ``dual '' conformal symmetry in  momentum space, studied recently in the context of \N=4 Yang-Mills theory in \cite{Drummond:2007cf} and associated with  string theory constructions in
\cite{Metsaev:1998it}-\cite{Claus:1999xr}.

\subsection{Manifestly supersymmetric local counterterms and  non-local effective actions}

A simple example would be to start with the 3-loop type invariant which is a square of a Bell-Robinson tensor. For the 4-particle case we are interested in the linearized part of the following expression
\be
S=\kappa^{4} \int d^4  x   C_{\alpha \beta \gamma \delta} (x)  C_{\dot \alpha \dot \beta \dot \gamma \dot  \delta}  (x) C^{\alpha \beta \gamma \delta}  (x)  C^{\dot \alpha \dot \beta \dot \gamma \dot  \delta}  (x) \ .
\ee
Here the dimension of the 4th power of the Weyl  curvature tensor is equal to 8, which is canceled by 4 integrations over $x$ and by $\kappa^{4}$ which adds another - 4. Since for the UV counterterms in L-loops one expects $\kappa^{2(L-1)}$, this explains the association of the square of the Bell-Robinson tensor, $(C_{\alpha \beta \gamma \delta}(x) C_{\dot \alpha \dot \beta \dot \gamma \dot  \delta}(x))^2$ with the 3d loop counterterm in supersymmetric theories \cite{Deser:1977nt},
\cite{Kallosh:1980fi}, \cite{Howe:1981xy}.
Before involving  supersymmetry we may just change our setting from the local integral to the box integral
\be
\kappa^{4} \int d^4 x  \quad  \Rightarrow \quad  \int {d^4x_1 d^4x_1  d^4x_1  d^4x_1 \over x_{12}^2 x_{23}^2  x_{34}^2  x_{41}^2 } \ ,
\ee
where $x_{12}= (x_1-x_2)$.
The negative dimension -8 comes now from the box integral, instead of the local $ \int d^4 x$, which requires  help from the gravitational coupling to balance the dimension of the $R^4$. Now we have to make sure that the product of the 4 curvatures taken at different points, $x_1, x_2, x_3, x_4$, is a scalar. For this we just have to work in a tangent space where all tensors are ``blind'' under the curved space transformation and transform only under the Lorentz transformation of the tangent space. Instead of $C_{\mu \nu \lambda \delta}(x)$ tensors which are contracted with $g^{\mu\nu}(x)= e_a^\mu(x ) e^{a\nu} (x) $, the $x$-dependent metric,  we have to use $C_{abcd}(x) \equiv
e_a^\mu(x)  e_a^\mu(x)   e_a^\mu(x)   e_a^\mu(x)  C_{\mu \nu \lambda \delta}(x) $, a tangent space tensor which is to be contracted with $\eta_{ab}$, the space-time independent metric. Spinors of general relativity live in tangent space anyway, so  the contraction proceeds via $\epsilon_{\alpha \beta}$ and again we can contract spinors at different points in space-time and make invariants \footnote{The non-locality would also require corrections such as Wilson loops for the local symmetries joining the split integration points. However, this would not affect the lowest order terms which we discuss. }.
This means that at the linear level the expression
\be
S=\int {d^4x_1 d^4x_2  d^4x_3  d^4x_4 \over x_{12}^2 x_{23}^2  x_{34}^2  x_{41}^2 } C_{\alpha \beta \gamma \delta}  (x_1)  C_{\dot \alpha \dot \beta \dot \gamma \dot  \delta}  (x_2) C^{\alpha \beta \gamma \delta}  (x_3)  C^{\dot \alpha \dot \beta \dot \gamma \dot  \delta}  (x_4)
\label{BR}\ee
may serve as a part of the 1-loop effective action for 4-particles and not only as a  3-loop counterterm. An additional term will come from exchanging $x_2 \Leftrightarrow x_3$.

In the  \N=8 case,    a linearized superinvariant, a candidate for the 3-loop counterterm,  is known:
\be
S_{{\cal N}=8}= \kappa^{4} \int d^4 x\ d^{16}  \theta_{B}\   W^4(x, \theta_B)  \ ,
\ee
where the index $B$ means either a specific choice of the basis in $\theta$-space as found in \cite{Kallosh:1980fi} or a representation $[232,848]$ of the $SU(8)$ used in \cite{Howe:1981xy} (a particular Young tableau), which makes the \N=8 supersymmetric invariant  also manifestly $SU(8)$ symmetric.

Now for our purpose to get a non-local 1-loop supersymmetric invariant,  we may look at  $\kappa$-independent terms with the box integral. For \N=8 supergravity it will be easy to use the already known N=1 supersymmetric answer and to generalize it to 8 supersymmetries. We  also look for other possible terms in the effective action.

\section{ \N=4 super-Yang-Mills theory}
As a warm-up we consider the effective action of the \N=4 super-Yang-Mills theory.
The on-shell action is known as a function of the on-shell Lie-algebra valued superfield
\be
W_{ij}= t^I W _{ij}^I = - W_{ji} \ , \qquad W_{ij} = {1\over 2} \epsilon_{ijkl}\bar W^{kl} \ ,
\ee
where $t^I$ is a generator of a non-Abelian group. The superfield satisfies constraints which are equivalent to equations of motion.
\be
\nabla_{\alpha i} W_{jk}= \nabla_{\alpha[i} W_{jk]} \ , \qquad \bar \nabla_{\dot \alpha}^ i W_{jk}= -{2\over 3} \delta^i{}_{[j } \nabla_{\dot \alpha}^l W_{k]l} \ .
\ee
We may slightly generalize the expression for the local quartic invariant presented in  \cite{Howe:1981xy} so that it can describe a 1-loop effective action of \N=4 super-Yang-Mills theory. Namely, we first introduce the covariant spinorial derivative action of the superfield
\be
\nabla_{\alpha}{}^{ i} W_{jk} = D_{\alpha}{}^{i}  W_{jk} + [A_\alpha^i, W_{jk}] = \delta^{i}_{[j }\lambda_{\alpha k]} \ .
\ee
The non-local 1-loop box-type effective action can be given in \N=4 manifestly supersymmetric form as follows
\be
  \int  d^4 x_1 d^4 x_2  d^4 x_3  d^4 x_4  D^{[ij],[kl] }\bar D^{[pq], [rs]}\,   {{ \rm Tr} \left (W_{ij}(x_1, \theta, \bar \theta) W_{kl}(x_2, \theta, \bar \theta) W_{pq}(x_3, \theta, \bar \theta) W_{rs}(x_4, \theta, \bar \theta)\right)_{\bf 105}\over x_{12}^2 x_{23}^2  x_{34}^2  x_{41}^2 } \ .
\label{action4}\ee
Here  $D^{[ij],[kl] }= (d\theta^{[i}_\alpha d\theta^{j]}_\beta  d\theta^{[k}_\alpha d\theta^{l]}_\beta)\,  \epsilon^ {\alpha \gamma} \epsilon^ {\beta \delta} $, and similar for $\bar D^{[pq], [rs]}$. The kernel (integrand) is in the representation  {\bf 105} of $SU(4)$.  For this representation it was proved in \cite{Howe:1981xy}  that the corresponding local quartic terms defined by the superaction with  integration over the incomplete superspace is manifestly \N=4 invariant. An analogous proof may be  valid for the non-local effective action in eq. (\ref{action4}).

The dimension of  the $x$-integration is -16 and that of the $\theta$-integration is +4, together  they give -12. The box part
  ${1\over x_{12}^2 x_{23}^2  x_{34}^2  x_{41}^2 }$,
  adds +8 and the superfields add +4;  the total dimension is zero.

  At the linear level this simplifies since we can use  arguments analogous to the ones used in \cite{Kallosh:1980fi} for the linearized \N=8 superfield. Namely,  one can show that there exists a basis in which every $x$ is shifted so that
\be
x' _{\alpha \dot  \alpha}= x_{\alpha \dot  \alpha}+i \sum_1^2 \theta_i \sigma_{\alpha \dot  \alpha} \bar \theta^i - i \sum_3^4 \bar \theta^j \sigma_{\alpha \dot  \alpha}  \theta_j
\label{basisYM}\ee
and in such a basis  $W_{12}= \overline W^{34} \equiv  t^I \Phi^I $ depends only on $\theta^1, \theta^2$ and $\bar \theta^3, \bar \theta^4$. Thus at the linear level the four-particle 1-loop effective action takes a simple manifestly \N=4 supersymmetric form
\be
S^{\rm box} _{{\cal N}=4} \sim  \int  d^4 x_1 d^4 x_2  d^4 x_3  d^4 x_4  d^8\theta_B \,   {{\rm Tr}  \left (\Phi _1 \Phi_2 \Phi_3 \Phi _4\right) \over x_{12}^2 x_{23}^2  x_{34}^2  x_{41}^2 } \ .
\label{action4linear}\ee
Here
\be
\Phi_a \equiv \Phi (x_a, \theta_B) \ ,\qquad  a=1,2,3,4 \ ,\qquad   \theta_B=(\theta^1, \theta^2, \bar \theta^3, \bar \theta^4)
\ee
It is nice that the box depends only on coordinate differences, like $x_{12}^2=(x_1-x_2)^2$ which is invariant under the change of the basis in eq. (\ref{basisYM}).
In this form the action is not manifestly $SU(4)$ symmetric as it is in (\ref{action4}). However, it is easy to show that in fact, it must be $SU(4)$ symmetric. The pure gluon part of this action is an $SU(4)$ invariant, therefore all supersymmetric  partners must also be  $SU(4)$ invariant.
\be
S^{\rm box}_{{\cal N}=4} \sim  \int  d^4 x_1 d^4 x_2  d^4 x_3  d^4 x_4  \;  { { \rm Tr } \; \left (F_{\alpha \beta} (x_1) F^{\alpha \beta} (x_2) \bar F_{\dot \alpha \dot \beta}  (x_3) \bar F^{\dot \alpha \dot \beta}   (x_4)  + \rm sym \right)\over x_{12}^2 x_{23}^2  x_{34}^2  x_{41}^2 } +...
\label{gluons4linear}\ee
Here ... stands for supersymmetric partners.

A nicer description of the \N=4 Yang Mills theory is available in harmonic superspace where all local  $x$-space  invariants have been classified,  \cite{Galperin:2001uw}, \cite{Hartwell:1994rp}. It would be interesting to use the harmonic superspace also for the non-local in $x$-space invariants.

For the one-particle irreducible (OPI) effective action we may try to construct a triangle and a bubble from  polynomials in  \N=4 supersymmetric superfields.
The background functional can depend only on superfield $\Phi $ of dimension $+1$ and its covariant derivatives owing to gauge invariance. So let us see if we can have triangles. The candidate will be
\be
S_{{\cal N}=4}^{\rm triangle} \sim \int {d^4 x_1 d^4 x_2  d^4x_3     }\; d^8 \theta_B\; {{\rm Tr }\;
P(\Phi, \partial \Phi )    \over x_{12}^2 x_{23}^2  x_{31}^2   }\ ,
\label {triangle} \ee
where $P$ means a polynomial of all higher powers of superfields and their  derivatives. The dimension of the triangle is  - 12+6,  together with +4 from the $\theta$ integral this  makes -2. It can be saturated by 2 superfields, however, we need at least 3 to have  momentum outgoing from each vertex of the triangle. Adding derivatives will only increase the imbalance of dimensions, so there are no triangles in the effective OPI action in \N=4 Yang Mills.

The same is true for the bubble
\be
S^{\rm bubble}_{{\cal N}=4} \sim \int  d^4 x_1 d^4 x_2     \, d^8 \theta_B \,  {{ \rm Tr}  P(\Phi, \partial \Phi )  \over x_{12}^2 x_{21}^2     } \ .
\label {bubble}
 \ee
Here the counting gives -8+4 from the $x$- and $\theta$-integrations, the bubble in the denominator adds +4, so only a  zero number of superfields is possible: no bubbles in the effective action which would have on-shell \N=4 supersymmetry.

 For the case of  \N =1 one can use a chiral integral with a smaller size $\theta$ space and a spinorial \N=1 on-shell superfield has dimension  3/2.
\be
S_{{\cal N}=1}^{\rm bubble} \sim \int {d^4 x_1 d^4 x_2        }\, d^2 \theta\,  {{\rm Tr }  [W_\alpha (x_1,\theta)  W^{ \alpha} (x_2,\theta)] \over x_{12}^2 x_{21}^2} +c.c.
\label {bubbleN1} \ee
The dimension of the $x$ and $\theta$ integral gives -8+1,  from the bubble we get +4 and from the superfields +3. For \N=2 there is a chiral superspace and a chiral superfield $W_{ij} $ of dimension 1.
\be
S_{{\cal N}=2}^{\rm bubble} \sim \int {d^4 x_1 d^4 x_2        }\, d^4 \theta\, {{\rm Tr}    [ W_{ij}(x_1,\theta) W^{ij}(x_2,\theta) ] \over x_{12}^2 x_{21}^2} +c.c.
\label {bubbleN2} \ee
The counting goes as -8+2+2+4=0.
Thus we have no problems making bubbles in  \N=1 and \N=2 super Yang Mills theories.

Before we switch to supergravity,  we would like to note the relation of  bubbles to axial anomalies. Bubbles in  \N=1 and \N=2 super Yang Mills theories consist of two terms, one chiral, and one anti-chiral. We may also form a difference between these two terms which is relevant to an axial anomaly.
\be
(\partial_\mu J^{\mu 5}) _{{\cal N}=1}^{\rm bubble} \sim \int {d^4 x_1 d^4 x_2       }\, d^2 \theta\,   {{\rm Tr } [ W_\alpha (x_1,\theta) W^{ \alpha}(x_2,\theta)]  \over x_{12}^2 x_{21}^2 } -c.c.
\label {bubbleN1} \ee
\be
(\partial_\mu J^{\mu 5}) _{{\cal N}=2}^{\rm bubble}  \sim \int {d^4 x_1 d^4 x_2        }\, d^4 \theta\,  {{\rm Tr  }   [W_{ij}(x_1,\theta) W^{ij}(x_2,\theta ]  \over x_{12}^2 x_{21}^2} - c.c.
\label {bubbleN2} \ee
If we would shrink  the bubbles in eqs.  (\ref{bubbleN1}), (\ref{bubbleN2}) (the dimension of $d^4 x $ is the same as ${d^4x_1 d^4 x_2\over x_{12}^2 x_{21}^2}$), we would get the supersymmetric generalization of the topological invariant  $ F_{ab}^*F^{ab}$. However, in  \N=4 case we do not have supersymmetric bubbles and there is no supersymmetric generalization of the  local $  F_{ab}^*F^{ab}$ term. This means that the contribution of fermions to 1-loop axial anomaly must cancel in \N=4.

No such terms can be associated with the box graph in eq.   (\ref{action4}) , (\ref{action4linear})  since each term is real. It is also interesting that if in the \N=2 theory we were to add matter, that removes the first loop divergence and axial anomaly, we also get   a UV finite theory, since \N=2 is UV finite starting from the second loop. This also depends on the non-renormalization theorem for the gauge-coupled hypermultiplet \cite{Howe:1983sr}.

If one tries to balance the dimension of the 1-loop n-gons with $n>1$ in \N=4 YM , it is possible, for example, to take an n-gon with n integral which gives -4n, 8 $\theta$ integrations with +4 and n superfields with 2d derivatives, or more than one superfield in a given  vertex.  For an example of  such a case with one superfield in each vertex,  one finds
$
-4n+4 +2n +n+2d=0 $ and $ n=4+2d$.
If we take total superspace integrals for the n-gon and one superfield in each vertex,
$
-4n+8 +2n +n+2d=0 $ and $ n=8 +2d
$.
An example would be an octagon with integration over the 16 $\theta$'s and with a product of 8 superfields, one in each vertex. From the integrations we  get -32+8, from the 8 superfields we get 8 and 16 from 8 propagators. Many such examples seem to be possible. However, they always require $n\geq 4$.

\section {\N=8 supergravity}
The linearized superfield of \N=8 supergravity is
\be
W_{ijkl} = {1\over 4!} \epsilon _{ijklmnpr} \overline W^{mnpr} \ .
\ee
  We will use here,  for simplicity,  the setting in \cite{Kallosh:1980fi} where
 the linear superfield $W_{1234} $  depends only on 16 $\theta$'s
 \be
 W_{1234} = \overline W^{5678} \equiv  W (x', \theta_B) \ , \qquad  \theta_B = (\theta_1, \theta_2, \theta_3,  \theta_4;  \bar \theta^5,\bar \theta^6, \bar \theta^7, \bar \theta^8)
 \ee
 in a special basis defined in  \cite{Kallosh:1980fi}, analogous to eq. (\ref{basis}) in YM.
 \be
x' _{\alpha \dot  \alpha}= x_{\alpha \dot  \alpha}+i \sum_1^4\theta_i \sigma_{\alpha \dot  \alpha} \bar \theta^i - i \sum_5^8 \bar \theta^j \sigma_{\alpha \dot  \alpha}  \theta_j \ .
\label{basis}\ee
This basis allows one to prove the linearized \N=8 supersymmetry of the candidate for the 3-loop counterterm. This setting was further developed in  \cite{Howe:1981xy} so that the $SU(8)$ symmetry became manifest. The two answers for the 3-loop counterterm are the same since the term quartic in the Weyl tensor in \cite{Kallosh:1980fi} is $SU(8)$ invariant, so the rest of the partners have to be $SU(8)$ invariant. Now we can use this superfield to make an \N=8 supersymmetric box.

 The \N=8 supersymmetric version of the 1-loop on-shell effective action is rather simple:
\be
 S^{\rm box}_{{\cal N}=8} \sim \int d^4 x_1 d^4 x_2  d^4x_3  d^4 x_4 \; d^{16} \theta_B \;  {W_1 W_2  W_3 W_4  \over x_{12}^2 x_{23}^2  x_{34}^2  x_{41}^2 } \ ,
\label{8} \ee
where $W_a\equiv W(x_a,\theta_B), a=1,2,3,4$. The counting of dimensions proceeds as follows: from the $x$  and $\theta$ integrations we get -16+8, from the 4 propagators, +8,  and zero from the superfields, since each $W$ has dimension zero. Eq.  (\ref{8})  provides  an \N=8 supersymmetric extension to the 1- loop box diagram for the square of the Bell-Robinson term with each Weyl spinor $C$ at a different corner of the box.  This means that  eq. (\ref{8})  after $\theta$-integration will have in addition to the
pure gravity part, as in eq.  (\ref{BR}), also a contribution from the 8 gravitinos and  all members of the \N=8 multiplet. It also provides an \N=8 supersymmetric generalization of the \N=1 supergraph result for \N=8, obtained in \cite{Grisaru:1981ye}.

The field $W$ is a singlet as distinct from a non-Abelian field in  (\ref{action4linear}),  therefore the symmetry of interchanges  the corners of the box  in (\ref{8}) does not involve the non-Abelian generators. The total answer in (\ref{8}) looks like a box with 4 real scalar fields. This explains easily why the double IR divergence, proportional to $s+t+u=0$ is not present  in \N=8 supergravity. In the \N=1 answer in \cite{Grisaru:1981ye} there is a term proportional to
$W_{\alpha\beta \gamma} (x_1) W^{\alpha \beta \gamma} (x_2)  \overline W^{\dot \alpha \dot \beta \dot \gamma} (x_3) W_{\dot \alpha \dot \beta \dot \gamma}(x_4) $ and another one proportional to ${1\over 2} W_{\alpha\beta \gamma} (x_1) W^{\alpha \beta \gamma} (x_3)  \overline W^{\dot \alpha \dot \beta \dot \gamma} (x_2) W_{\dot \alpha \dot \beta \dot \gamma}(x_4) $. The IR divergence cancels between these two terms, but it is present in each of them. Thus, in the box-symmetric  \N=8 form of the answer in (\ref{8}) there is a reflection of the fact that  gravity has a softer IR divergence than Yang Mills theory.
In the YM case (\ref{action4linear}) on the other hand the symmetry between corners of the box is more complicated owing to an extra permutation of non-Abelian generators in each leg,
$
{\rm Tr} (t^I t^J t^K t^L) \, (\Phi_1^I \Phi_2^J \Phi_3^K \Phi_4^L)
$.

One can look now for triangles and bubbles as we did in \N=4 YM. The counting of dimensions proceeds as follows: from the $x$  and $\theta$ integrations we get -12+8, from the 3 propagators we get  +6 and we get zero from the superfields, since each $W$ has dimension zero. The total is +2, the derivatives will only increase the dimensions and we look only for the OPI effective action, so no other sources of negative dimensions are possible: any term of the form $(p_1\cdot p_2)^{-1} $ would give a pole, which should not be present  in the background OPI action. For the bubbles we have -8+8 from the integrations, from the  2 propagators we get 4 and only non-negative contributions from superfields and derivatives. So we may try  arbitrary polynomials in superfields and its derivatives to construct triangle and bubbles:
\be
  \int {d^4 x_1 d^4 x_2 d^4 x_3       }\, d^{16} \theta_B \, {P [  W , \partial W ]_B  \over x_{12}^2 x_{23}^2 x_{31}^2}\ , \qquad  \int {d^4 x_1 d^4 x_2        }\, d^{16} \theta\, {P [  W , \partial W ] _B \over x_{12}^2 x_{21}^2 } \ .
\label {BT8} \ee
These do not exist in the first loop with $\kappa^2(L-1)$ since the superfield dimension is zero, the derivatives in space-time as well as in $\theta$ only increase the dimension,  and we get $-6+8 +d\neq 0$ for  triangle with any number of superfields at each  vertex and $-4+ 8 +d \neq 0$ for a bubble with any number of superfields at each  vertex. Here $d$ is a positive number of   the derivatives in $P$.

It would be interesting to perform an analysis analogous to the one in this paper using the  linearized form of the harmonic superspace of \N=8 supergravity, discussed in \cite{Hartwell:1994rp}.

Before leaving this subject,  we may ask what is the analogous situation in \N$ <8$. The  \N=4 supergravity case is easy since the linear superfield is a chiral scalar of weight 0, $\nabla_{\dot \alpha} ^i W=0, i=1,2,3,4$. So one can build  \N=4 supersymmetric bubbles
\be
S_{{\cal N} =4}^{\rm bubble} \sim \int {d^4 x_1 d^4 x_2}\, d^8 \theta\, { W^n(x_1, \theta) W^m(x_2, \theta) \over x_{12}^2 x_{21}^2 } + c.c.
\label {bubble4} \ee
These are possible since the dimension of the chiral scalar  \N=4 superfield $W$ is zero and \\
-4+4=0 and we have $n$ superfields in one vertex and $m$ in the other.
 For the triangle we can allow any number of superfields in each vertex and take two derivatives and use  -12+6+4 +2=0
 \be
S_{{\cal N}= 4}^{\rm triangle} \sim \int {d^4 x_1 d^4 x_2  d^4x_3     }\, d^8 \theta\, { W^m(x_1) \partial W^n(x_2) \partial W^l(x_3) \over x_{12}^2 x_{23}^2  x_{31}^2} +c.c.
\label {triangle 4} \ee
From the available bubbles and triangles in \N=4 supergravity we immediately deduce that there are possible candidates for axial anomalies
\be
(\partial_\mu J^{\mu 5}) _{{\cal N}=4}^{\rm bubble} \sim \int {d^4 x_1 d^4 x_2         }\, d^8 \theta\, { W^n(x_1, \theta) W^m(x_2, \theta) \over x_{12}^2 x_{21}^2 } - c.c.
\label {bubbleN4}
\ee
and from the triangle
\be
(\partial_\mu J^{\mu 5}) _{{\cal N}=4}^{\rm triangle} \sim \int {d^4 x_1 d^4 x_2  d^4x_3     }\, d^8 \theta\, { W^m(x_1) \partial W^n(x_2) \partial W^l(x_3) \over x_{12}^2 x_{23}^2  x_{31}^2} -c.c.
\label {triangleN4} \ee
If we would shrink the bubble (\ref{bubbleN4}) in \N=4 supergravity to the point (same dimension) we would get for $m=n=1$  a supersymmetric generalization of the topological invariant
$C_{abcd}^*C^{abcd}$ as shown in \cite{Kallosh:1980fi}. No such terms are available 
 in \N=8.  So in \N=8 the contribution from all fermions to the 1-st loop triangle anomaly should cancel.

Our analysis of the absence of the OPI triangle and bubble graphs in the background functional with manifest \N=8 supersymmetry is related to the no-triangle hypothesis \cite{BjerrumBohr:2006yw}  in  \N=8 supergravity. Hopefully, this relation can be clarified and our dimensional superspace analysis can be used for a  confirmation of this hypothesis.

According to \cite{Bern:2007hh}, \N=8 supergravity is  3 loop finite. On the other hand,  the structure of the 1-loop effective action in eq. (\ref{8}) is closely related to the local counterterm for the 3-loop UV divergence.  This is somewhat surprising and may need an explanation. One possible explanation is given by the non-renormalization theorem in \cite{Howe:2002ui}, however, it comes with the prediction that \N=8 supergravity may be UV divergent starting from $L=5$, so some other explanations may also be possible.
If we understand this, perhaps we will know more about the conjectured all-loop finiteness of \N=8 supergravity \cite{Bern:2006kd}.

It is also interesting to use our setup for the \N=0 case. The relevant 1-loop triangle and the bubble are:
 \be
S_{{\cal N}=0}^{\rm triangle} \sim \int {d^4 x_1 d^4 x_2  d^4 x_3   \over x_{12}^2 x_{23}^2  x_{31}^2   }\, C^{abcd} (x_1) C_{cd}{}^{ef}  (x_2) C_{efab}(x_3) \ ,
\label {triangle0}
\ee
\be
  S_{{\cal N}=0}^{\rm bubble } \sim \int {d^4 x_1 d^4 x_2    \over x_{12}^2 x_{21}^2   }\, C^{abcd} (x_1) C_{abcd}  (x_2) \ .
\label {bubble0} \ee
The 1-loop finiteness of pure gravity can be seen here as follows. The local counterterm corresponding to a bubble which shrinks to a point vanishes since the local terms quadratic in the Weyl tensor vanishes on-shell due to the Gauss-Bonnet identity, whereas the bubble does not vanish since each $C$ is at a different point. Note also that there are  n-leg triangle graphs here when each of the curvatures is expanded  in powers of metric as well as when the tree replaces a single graviton line. Such n-leg triangles  are allowed by dimensionality: apparently they source the 2-loop UV infinity in the second loop in pure gravity.

\section{On the dual conformal symmetry of the effective action}

Here we will present our results for non-local manifestly supersymmetric \N=4 SYM and \N=8 supergravity effective actions using dual conformal coordinates \cite{Drummond:2007cf}.
Consider the box part of the \N=4 Yang Mills effective action  (\ref{action4linear}) in  Fourier space with $p_1+p_2+p_3= p_4$,
\begin{eqnarray}
\int {d^4 k \over k^2 (k-p_1)^2  (k-p_1-p_2)^2  (k+p_4)^2 }\, d^8 \theta_B \,  {\rm Tr }
 \left[ \Phi (p_1, \theta_B) \Phi  (p_2, \theta_B) \Phi (p_3, \theta_B)  \Phi (p_4, \theta_B) \right].
\label {4momentum} \end{eqnarray}
Following \cite{Drummond:2007cf}, we will introduce the ``coordinates''  $y_{jk}= y_j- y_k$ which are dual to the usual  coordinates and are expressed via momenta in the box integrals
\be
p_1=  y_{12}\ , \quad   p_2=  y_{23}\ , \quad p_3=  y_{34}\ , \quad p_4=  y_{41}\ , \quad k=y_{15} \ .
\label{dual} \ee
These dual coordinates provide a spectacular dual conformal symmetry of the box integral  \cite{Drummond:2007cf}.
In terms of the new variables  we find the following answer for the box part for the one-loop action in \N=4 SYM:
\be
 S^{\rm  dual box}_{{\cal N}=4}\sim  \int {d^4 y_5  \over y_{15} ^2 y_{25}^2  y_{35}^2  y_{45}^2 }\, d^8 \theta_B \, {\rm Tr }\;  \left[ \Phi (y_{12} , \theta_B) \Phi (y_{23} , \theta_B)  \Phi (y_{34} , \theta_B)  \Phi (y_{41} , \theta_B) \right].
\label {emery} \ee

We find an analogous expression  for \N=8 supergravity,  starting from the Fourier transform of the expression in  eq. (\ref{8}) and using the dual coordinates in eq. (\ref{dual}):
 \be
 S^{\rm dual box}_{{\cal N}=8} \sim \int {d^4 y_5  \over y_{15} ^2 y_{25}^2  y_{35}^2  y_{45}^2 } \; d^{16} \theta_B \; \left[W (y_{12} , \theta_B)W (y_{23} , \theta_B)  W (y_{34} , \theta_B)  W(y_{41} , \theta_B) \right].
\label{dual8} \ee

 It is rather interesting to look at  the expression for the bosonic part of the action  after integrating over $\theta$. In the \N=4 SYM case we find
\be
 \int {d^4 y_5  \over y_{15} ^2 y_{25}^2  y_{35}^2  y_{45}^2 }\,  {\rm Tr} [ F_{\alpha\beta} (y_{12} ) F^{\alpha\beta} (y_{23} ) F_{\dot \alpha \dot \beta } (y_{34}) F^{\dot \alpha \dot \beta}  (y_{41} ] +  ...
\label {renata1} \ee
Here $ F_{\alpha\beta}$ and $F_{\dot \alpha \dot \beta }$ are the vector field strength spinors.
In \N=8 supergravity we find
\be
 \int {d^4 y_5  \over y_{15} ^2 y_{25}^2  y_{35}^2  y_{45}^2 }\,  C_{\alpha\beta\gamma\delta} (y_{12} ) C^{\alpha\beta\gamma\delta} (y_{23} ) C_{\dot \alpha \dot \beta\dot \gamma\dot \delta} (y_{34}) C^{\dot \alpha \dot \beta\dot \gamma\dot \delta}  (y_{41} )+  ...
\label {renata2} \ee
Here  $C_{\alpha\beta\gamma\delta}$ and $C_{\dot \alpha \dot \beta\dot \gamma\dot \delta} $ are conformal Weyl spinors.
These equations show  an intriguing relation  to (super)-conformal symmetry in the dual coordinates related to the momenta of particles. These dual coordinates are also related to the string action.

The quantization of the Green-Schwarz action in an $AdS_5\times S^5$ background \cite{Metsaev:1998it} in Killing covariant background gauges was performed in \cite{Kallosh:1998nx}. This action acquires a simple form after a dualization as performed
 in \cite{Kallosh:1998ji}. These dual coordinates of the GS string action have been   used  in
\cite{Alday:2007hr} in studies of gluon amplitudes. A detailed investigation of this duality and the integrability of string theory in an AdS background was recently performed in \cite{Ricci:2007eq}.

It is also quite interesting that these dual string coordinates have been   related to twistors and supertwistors in \cite{Claus:1999xr}. In particular, the BRST quantization of the particle in $AdS_5$ was performed on the basis of the dual coordinates which were replaced by twistors. It would be interesting to understand the dual conformal symmetries   of \N=8 supergravity  better and to find out if they play any role in the UV properties of the four-dimensional theory.

The story of magic identities \cite{Drummond:2007cf} becomes even more interesting when one learns that the dual conformal symmetry was checked through five loops  in explicit computations in \cite{Bern:2006ew}.

\section{Discussion }

The main result of this paper is an expression (\ref{8}) for the \N=8 supergravity 1-loop on-shell effective box-type action in the background field method, which has a manifest \N=8 supersymmetry at the linear level.  Our result suggests  that the local 3-loop UV counterterm $  (R_{....})^4$ constructed for \N=8 supergravity in \cite{Kallosh:1980fi,Howe:1981xy}  does not show up not because we do not know how to make it non-linear. Indeed, it does show up in the non-local expression in the 1-st loop in eq.  (\ref{8}).  So something else prevents it from appearing as a UV divergence.

What can we deduce from this result about the 5-loop counterterm of \cite{Howe:2002ui} or the  8-loop and higher loop counterterms of \cite{Howe:1980th}, \cite{Kallosh:1980fi}? No firm statements can be made at this time, but the lore that we should expect local counterterms seems to be less reliable than it was for a long time.

Apparently, the conjectured n-loop finiteness as well as  the 3-loop finiteness stated in \cite{Bern:2007hh} have two important features. The first feature is the unitarity cut rules.
These rules have been tested in various special situations. On the other hand, the standard rules are based on  the path integral quantization and lead to virtual lines in Feynman graphs. Path integral quantization with proper ghost actions is the basis for the standard proof that the theory is unitary; these methods are well established and do not require testing. However, computationally, the unitarity cut rules, which allow one to construct higher order diagrams  in terms of lower order on-shell amplitudes,  are more efficient for investigation of the divergences and for constructing  finite answers. It would be remarkable if they could be derived for all possible sets of graphs via some type of path integral  method which would keep  combinatorial properties  of all graphs under control.
The second feature is  the no-triangle property  of the 1-loop graphs in \N=8 supergravity  \cite{BjerrumBohr:2006yw}.  For the use in generic
higher loop graphs  it remains a conjecture. For  the  3-loop explicit
calculation it only served as a  motivation to do the
calculation.  Therefore one may expect that some of
the higher loop cancellations can be understood from the no-triangle
property of the 1-loop graphs.

In this paper we used superspace dimensional analysis  and demonstrated that no triangle and bubble graphs for any number of external legs are possible  in the OPI effective action with manifestly realized \N=8  supersymmetry. To the best of our understanding, this fact corresponds to the no-triangle hypothesis  for the 1-loop graphs in \N=8 supergravity as formulated in  \cite{BjerrumBohr:2006yw},  despite the difference in the
 two approaches. If the relation between our approach and that in  \cite{Bern:2007hh}  can be fully clarified, it may lead to a proof of the  no-triangle property  of the 1-loop graphs in \N=8 supergravity in the form required in \cite{BjerrumBohr:2006yw}.
 It would mean that at least one ingredient of the package for all-loop finiteness  in \cite{Bern:2006kd} is firmly established  from the supergravity point of view.

We have  also shown that there is a  1-loop ``bad triangle'' in the \N=0 case, see eq. (\ref{triangle0}), and that there is a ``bad bubble,'' see eq. (\ref{bubble0}),  so according to the logic in \cite{Bern:2007hh} there is no reason to expect an absence of the 2-loop divergence in pure gravity using their methods of computation. In other words, there is no conflict between the possibility of the finiteness of the N= 8 supergravity and the well-known fact of the existence of the 2-loop divergence in pure gravity.

In this paper we have shown that there are no 1-loop \N=8 supersymmetric generalizations of the \N=0 triangles  (\ref{triangle0}) and bubbles (\ref{bubble0}) in the effective action, but they do exist for smaller \N$<8$.  Moreover, we argued that the presence of a scalar box and the absence of triangles and bubbles  in the finite part of  \N=8 1-loop effective action  is related to the absence of 1-loop conformal and axial anomalies. One may speculate that   the axial anomaly is a  1-loop phenomenon, and the conformal anomaly should follow the same rule due to supersymmetry. In such case, it would not seem implausible that  \N=8 supergravity  may be anomaly-free and UV finite. In conclusion, we believe that  better  understanding of quantum \N=8
supergravity,  as suggested  in \cite{Bern:2007hh,Bern:2006kd} and as analyzed in the superspace formulation of this paper,  deserves  full attention  and requires more studies.

\section*{Acknowledgments}

I am very grateful to Z. Bern,  L. Dixon,  S. Ferrara,  M. Grisaru and K. Stelle for helping me to understand and clarify the issues discussed in this paper. I thank
P. Howe, A. Linde,  U. Lindstrom,  J. Maldacena, S. Shenker,  M. Spradlin   and L. Susskind  for most useful discussions of \N=8 supergravity and \N=4 SYM theory.  This work is
supported by the NSF grant 0244728.


\end{document}